\documentclass[%
superscriptaddress,
preprint,
 amsmath,amssymb,
 aps,
]{revtex4-1}

\usepackage{graphicx}% Include figure files
\usepackage{dcolumn}% Align table columns on decimal point
\usepackage{bm}% bold math
\usepackage{epstopdf}

\usepackage{amssymb}
\usepackage{amsmath}

\newcommand{\mm}{\mathrm}
\newcommand{\ud}{\mm{d}}

\newcommand{\bi}{\begin{itemize}}
\newcommand{\ei}{\end{itemize}}

\def\T{\intercal}

\usepackage{xcolor}

\def\sign{\mathop{\sf sign}}

\def\ker{\mathop{\sf ker}}

\begin{document}

%\preprint{APS/123-QED}

\title{Fundamental limitations of network reconstruction}% Force line breaks with \\

\author{Marco Tulio Angulo}
\affiliation{%
 Center for Complex Networks Research, Northeastern University, Boston MA 02115, USA
}%
\affiliation{Channing Division of Network Medicine, Brigham and Women's
Hospital, and Harvard Medical School, Boston MA 02115, USA}


\author{Jaime A. Moreno}
 \affiliation{
Instituto de Ingenier\'ia, Universidad Nacional Aut\'onoma de M\'exico, Distrito Federal 04510,  M\'exico
}%

\author{Gabor Lippner}
\affiliation{%
 Department of Mathematics, Northeastern University, Boston MA 02115, USA
}%

\author{Albert-L\'{a}szl\'{o} Barab\'{a}si$^{1,}$}
\affiliation{%
Center for Cancer Systems Biology, Dana-Farber Cancer Institute,
Boston MA 02115, USA
}%
\affiliation{
Center for Network Science, Central European University, Budapest 1052, Hungary
}

\author{Yang-Yu Liu$^{2,5,*}$}


\date{\today}% It is always \today, today,

\begin{abstract}
Network reconstruction is the first step towards understanding, diagnosing and controlling the dynamics of complex networked systems. It allows us to infer properties of the interaction matrix, which characterizes how nodes in a system directly interact with each other. Despite a decade of extensive studies, network reconstruction remains an outstanding challenge. The fundamental limitations governing which properties of the interaction matrix (e.g., adjacency pattern, sign pattern and degree sequence) can be inferred from given temporal data of individual nodes remain unknown. 
Here we rigorously derive necessary conditions to reconstruct any property of the interaction matrix. These conditions characterize how uncertain can we be about the coupling functions that characterize the interactions between nodes, and how informative does the measured temporal data need to be; rendering two classes of fundamental limitations of network reconstruction. 
Counterintuitively, we find that reconstructing any property of the interaction
matrix is  generically as difficult as reconstructing the interaction matrix itself, requiring equally informative
temporal data.
Revealing these fundamental limitations shed light on the design of better network reconstruction algorithms, which offer practical improvements over existing methods.

\end{abstract}

\pacs{Valid PACS appear here}% PACS, the Physics and Astronomy
\maketitle

%\tableofcontents

%%%%%%%%%%%%%%%%%%%%%%%%
%%%%
%%%%
%%%%
%%%%
%%%%
%%%%%%%%%%%%%%%%%%%%%%%%

Networks are central to the functionality of complex systems in a wide range of fields, from physics to engineering, biology and medicine ~\cite{Barabasi:15,Newman:03,Dorogovtsev-RMP-08}. 
%~\cite{albert2002statistical,Newman:03,Dorogovtsev-RMP-08}. 
%.
%When a network serves as conduit to the system dynamics, it not only reveals the interconenction architecture of complex systm, but it also fundamentally affects the behavior.
%
When these networks serve as conduit to the system dynamics, their properties  fundamentally affect the dynamic behavior of the associated system; examples include   epidemic spreading~\cite{Pastor-Satorras-PRL-01,Cohen-PRL-00},
synchronization phenomena~\cite{Nishikawa-PRL-03,Wang-BC-05}, controllability~\cite{liu2011controllability,nepusz2012controlling} and observability~\cite{liu2013observability}.
For many complex networked systems, measuring the temporal response of individual nodes (such as proteins, genes and neurons)  is becoming more accessible \cite{Timme:14}.
Yet, the network reconstruction (NR) problem ---that is, recovering the underlying interconnection network of the system from temporal data of its nodes--- remains a challenge \cite{Timme:14, Villaverde:14,Prill:10}.
Consider a networked system of $n$ nodes. Each node is associated with a state variable $x_i(t) \in \mathbb{R}$, $i=1,\cdots,n$, at time $t$ that may represent the concentration of certain biomolecule in a biochemical system, the abundance of certain species in an ecological system, etc.  The time evolution of the state variables is governed by a set of ordinary differential equations:  
\begin{equation}
\label{system}
% \frac{ \ud x_i(t) }{\ud t}
 \dot x_i(t) = \sum_{j=1}^n a_{ij} f_{ij} \left(x_i(t), x_j(t) \right) + u_i(t), \quad i=1, \cdots, n.  
 \end{equation}
%Let's explain all terms. 
Here the \emph{coupling functions}  $f_{ij} :
\mathbb{R} \times \mathbb{R} \to \mathbb{R}$ 
specify the interactions between nodes ---self interactions when $i=j$, or pairwise interactions between nodes when $i \neq j$.
%
%Our results also apply to models with arbitrary interactions and known inputs, see SI-X.
%
The term $u_i(t) \in \mathbb{R}$ represents known signals or control
inputs that can influence the $i$-th state variable. 
The \emph{interaction matrix} $A = (a_{ij}) \in \mathbb R^{n \times n}$ captures the direct interactions between nodes, naturally defining the  interconnection network of the system by associating $a_{ij}$ to the link $j \rightarrow i$ between node $i$ and node $j$. 
 By appropriately choosing   the coupling functions, Eq.  \eqref{system} can model a broad class of networked systems  \cite{barzel2013universality}.
Given some function $\mathcal P$ of the interaction matrix ---which we call a \emph{property}--- 
% Given some \emph{property} $\mathcal P: \mathbb R^{n\times n} \rightarrow  \mathbb R^{n\times n}$,  
 NR aims to recover  the value  of $\mathcal P(A)$ from given temporal \emph{data} $\{x_i(t), u_i(t) \}_{i=1}^n$, $\forall t \in [t_0, t_1]$, and given \emph{uncertainty} of the coupling functions.% (such uncertainty is made precise later on).
 %
%Our analysis extends straightforwardly to systems with more general interactions, SI-12.

Note that the classical parameter identification (PI) problem for \eqref{system}  aims to recover the interaction matrix itself  (i.e. reconstructing the identity property) \cite{walter1997identification,ljung1998system,kailath2000linear}. 
But in many cases, instead of reconstructing %the link-weight matrix 
$A$ itself, we may want to reconstruct properties like
its \emph{sign pattern} $S= [s_{ij}] = [\sign(a_{ij})] \in \{-1,0,1\}^{n
  \times n}$, \emph{connectivity pattern} $C= [c_{ij}] =
[|s_{ij}|] \in \{0,1\}^{n \times n}$, \emph{adjacency pattern} $K=[k_{ij}] =[c_{ij} (1-\delta_{ij})] \in \{0,1\}^{n \times n}$ %with $k_{ij}=c_{ij}$ if 
($\delta_{ij}$ is the Kronecker delta)
%$i \neq j$ and 0 otherwise, 
or \emph{in-degree sequence} $\bm d=[d_i]=[\sum_j c_{ij}] \in \mathbb Z^n$. 
% 
%Here $\delta_{ij}$ is the Kronecker delta function.
%
Indeed, a key insight of network science is that important properties
of networked systems ---such as sign-stability,  structural controllability/observability and epidemic thresholds--- can be determined from $S$, $C$, $K$ or $\bm d$ without knowing $A$
~\cite{may1973qualitative,Cohen-PRL-00,Pastor-Satorras-PRL-01,Nishikawa-PRL-03,Wang-BC-05,shinar2010structural,liu2011controllability,liu2013observability,Ruths-Science-14}.
Note  that  these properties cannot be easily reconstructed by computing correlations in the data, simply because correlations capture both direct and indirect interactions.%--- despite networks are often constructed using such methods.
%
%\textcolor{red}{Key to the widespread use of network reconstruction has been the intuition that recovering less information should require less informative data. For example,   less informative data should be necessary to reconstruct $K$ than to reconstruct $A$ itself.}

NR helps us understand, diagnose and control the dynamics of diverse complex networked  systems, deepening our understanding of human diseases and ecological networks,  and letting us build more resilient power grids and  sensor networks
~\cite{wang2004visualization,arianos2009power,stein2013ecological,bonneau2008learning,tyson2001network,kohanski2010antibiotics,1509.06926}.
Yet, despite
% NR: difficulty and our findings
 a decade of extensive studies, NR remains
an outstanding challenge~\cite{Timme:14,Villaverde:14,filosi2014stability}. Many
existing algorithms do not perform significantly better than random
guesses~\cite{Prill:10, Villaverde:14}, and even well-established
methods %(driving-response experiments)  
can provide contradictory results for relatively simple
networks~\cite{prabakaran2014paradoxical}. 
It has been realized that these problems originate from our ignorance of the fundamental limitations of network reconstruction, governing which properties of the interaction matrix can be recovered from given temporal data and knowledge of the coupling functions \cite{Timme:14,Villaverde:14}.
Indeed, it is still unclear if an NR algorithm fails to recover the correct value for $\mathcal P(A)$ due to some design flaws, or due to limitations intrinsic to the available temporal data and/or our uncertainty about the coupling functions. % ---that is, due to non-identifiability implying that no algorithm can recover $\mathcal P(A)$. 
Furthermore, it is also unclear if NR can be solved with less informative data that that is necessary to solve the classical PI problem.
Our intuition suggests that NR is easier (in the sense of requiring less informative temporal data) than PI simply because we are recovering less information (e.g. $K$ instead of $A$). But, is this true?

Here we characterize the fundamental limitations of NR for the first time, by deriving necessary (and in some cases sufficient) conditions
to reconstruct any desired property of the interaction matrix.
We find that fundamental limitations arise from our uncertainty about the coupling functions, or uninformative  temporal data, or both. 
The first class of fundamental limitations is due to our uncertainty about the coupling functions, rendering a natural trade-off: the more information we want to reconstruct about the interaction matrix the more certain we need to be about the coupling functions. 
To show this,  we characterize necessary conditions that our uncertainty about the coupling functions needs to satisfy in order to reconstruct some desired property of the interaction matrix.  
For example, we show that it is possible to reconstruct the adjacency pattern $K$ without knowing exactly the coupling functions. But, in order to reconstruct the interaction matrix $A$ itself,  it is necessary to know these functions exactly.
Hence, if we are uncertain about the coupling functions, NR is easier than PI.

The second  class of fundamental limitations originates from uninformative data only, 
%from the measured temporal data, 
leading to a rather counterintuitive result: regardless of how much information we aim to reconstruct (e.g.  edge-weights,  sign pattern, adjacency pattern or in-degree sequence), the measured data needs to be equally informative.
This happens even if we know the coupling functions exactly.
We prove  that the same condition \eqref{PE} on the measured data is generically necessary regardless of the property to be reconstructed.
Hence, in the sense of informativeness of the measured data, reconstructing any property of the interaction matrix is as difficult as reconstructing the interaction matrix itself, i.e. NR is as difficult as PI.
% 
%not imply we can use data with less information. Indeed, regardless of the property to reconstruct and uncertainty in the system dynamics, the same (persistent excitation) condition \eqref{PE} is generically necessary.
%
%When only steady-state data from a single experiment is available,  our result implies that  generically no property of the interaction matrix can be reconstructed, not even the interaction matrix itself  \cite{sontag2002differential}.
%For example, this implies that it is generically impossible to reconstruct any property of the interaction matrix using steady-state data only.
%
In order to circumvent this limitation without acquiring more temporal data (i.e. performing more experiments), we show that prior knowledge of the interaction matrix is extremely useful.%, allowing to reconstruct 

%The two classes of fundamental limitations show that NR is useful when we are uncertain of the system dynamics, which happens in most complex systems. Yet, ironically, NR does not allow us to do more with less informative data ---not even if we are completely certain about the system dynamics.
These two classes of fundamental limitations indicate that when we are uncertain about the coupling functions (true for many complex systems) PI is impossible, but we can still reconstruct some properties of the interaction matrix $A$ provided the measured temporal data is  informative enough  and interactions are pairwise. 
In this sense, NR is easier than PI. 
Yet, ironically, even if we are completely certain about the coupling functions, with less informative data NR does not allow us to do more ---it is as difficult as PI.

\section{Results}

A property $\mathcal P(A)$ 
can be reconstructed if and only if (iff) %different $A$'s 
any two interaction matrices $A_1, A_2 \in \mathbb R^{n \times n}$ with different properties $\mathcal P(A_1) \neq \mathcal P(A_2)$ produce different node trajectories $\{x_i(t) \}_{i=1}^n, t \in [t_0, t_1]$, a  notion of \emph{identifiability} or
 \emph{distinguishability}   \cite{walter1997identification}.%, see
%Fig.~\ref{fig:distinguishability-network}. 
%
%Otherwise, it is impossible to reconstruct the desired property simply because we cannot decide if an interaction matrix with property $\mathcal P(A_1)$ or with property $\mathcal P(A_2)$ produced the observed node trajectories.
%
%The existence of \emph{indistinguishable} interaction matrices ---interaction matrices that produce identical node trajectories--- %whose properties have different values  
%with different properties establish the fundamental limitations of network reconstruction.

\begin{figure*}[!t]
\centering
\includegraphics[width=5.8in]{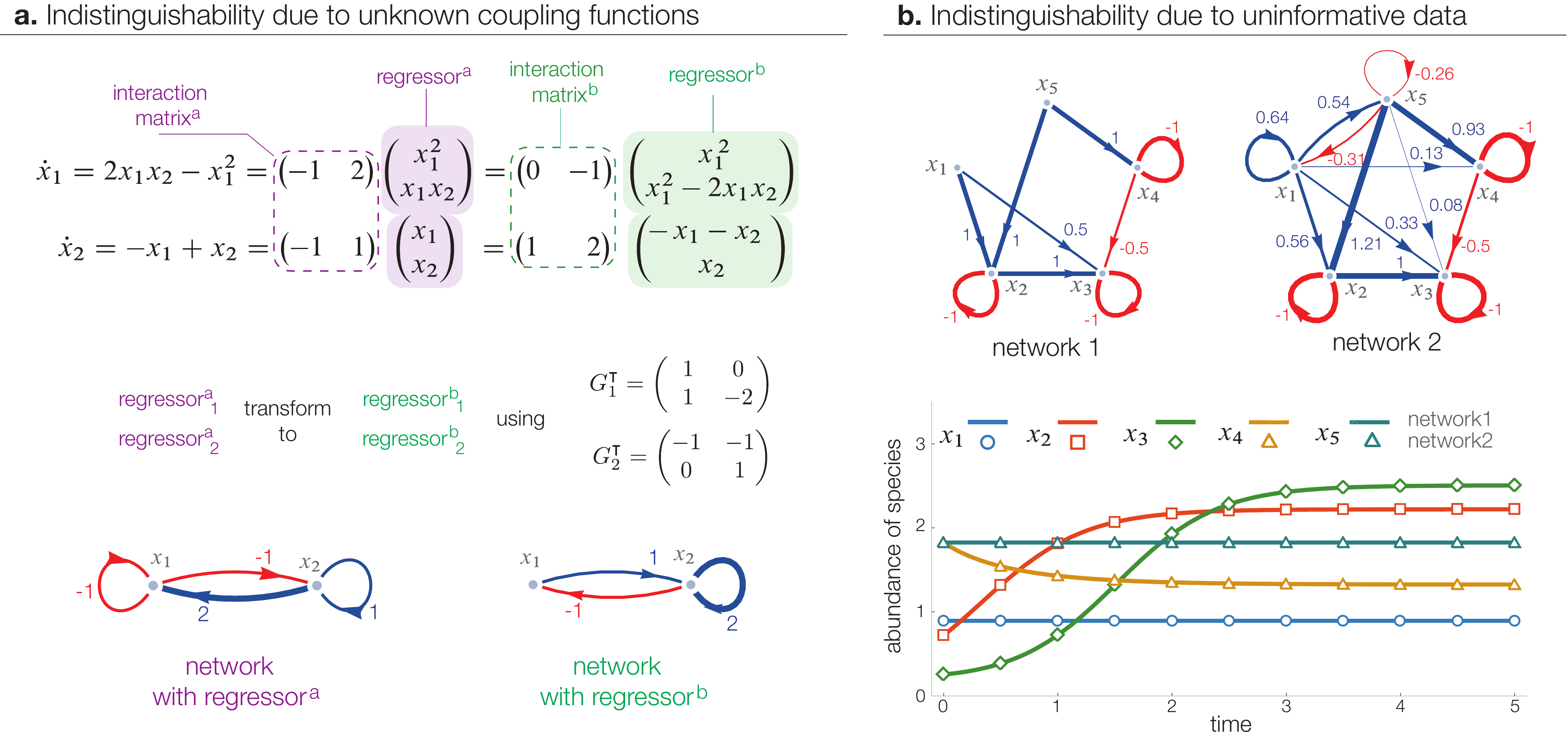}
\vspace*{-0.0cm}
\caption{\small 
% {\bf Indistinguishable networks prevent network
%     reconstruction.}
{\bf %Networks producing identical node trajectories are  indistinguishable.
Two sources of indistinguishability.
  } 
{\bf a.} 
The same dynamics can be characterized by two  regressors with different coupling functions (purple and green), yielding  indistinguishable networks that differ in their edge-weights, sign patterns, connectivity  patterns and degree sequences.
%
%Indeed, for any any nonsingular matrix $P \in \mathbb R^{n \times n}$,  $\dot x_i = \bm f_i \bm a_i^\T = \bm{\bar f_i} \bar {\bm a}_i^\T$ with $\bm{\bar f_i} = \bm f_i P $ and $\bar {\bm a}_i = \bm a_i P^{- \T}$.
%{\bf b.}
%%Two systems with different dynamics and networks  can generate 
%%identical node trajectories, because their dynamics can be written in different basis.
%%
%In this example, the energy $E(x_1, x_2) =  x_1^2 + x_2^2 = 1$ is conserved. Thus,  indistinguishable networks emerge because we can write $\dot x_3 = -x_3$ (with dynamics $f_{33} =x_3$) or equivalently $\dot x_3 = - (x_1^2 + x_2^2) x_3$  (with dynamics $f_{13} = x_1^2 x_3$ and $f_{23} = x_2^2 x_3$). 
%%
%This can also be interpreted as  rewriting the constant $1= x_1^2 + x_2^2$ using $x_1^2$ and $x_2^2$ as basis functions.
{\bf b.}
With the classical population dynamics described by the generalized Lotka-Volterra (GLV) model $\dot x_i = r_i x_i + \sum_j a_{ij} x_i x_j$, the two  different networks shown in the top panel
  ---representing two different inter-species interaction %community 
  matrices--- produce identical node trajectories $\bm x(t)$ (bottom panel). Here the growth rate vector is $\bm r=(0,-0.5, 0.5, -0.5, 0)^\T$ and initial abundance $\bm x(0) =(0.895349, 0.72093, 0.255814, 1.82558, 1.82558)^\T$.
  %
  %Here the node dynamics is given by the GLV model $\dot x_i = r_i x_i +
%\sum_j a_{ij} x_i x_j$  that has the form of Eq. \eqref{system} taking
%$f_{ij} = x_i x_j$ and $u_i(t)= r_i x_i(t)$, with $r_i$ a known
%species growth rate. % See Fig.\ref{SI-fig:distinguishability-networkSI} for additional details.
  %  Therefore, it 
 In these two examples,  it is impossible to reconstruct the edge-weights, sign-pattern, connectivity-pattern or degree sequence of the network
 simply because we cannot decide which one of the two networks
  produced the measured node trajectories.
%  
%
%reaction rate. 
%
%{\bf b.} According to \eqref{indistinguishable-vectors}, the
%indistinguishable interconnection vectors for each node are
%parametrized by $\ker M_i(t_0, t_1)$ with $t_0=0, t_1=5$. Since in
%this example $\ker M_i$ is not trivial, the regressor does not
%have Persistent Excitation and there will be nontrivial
%indistinguishable interconnection vectors. Indeed, network 2 in panel
%a. is obtained by adding these vectors with coefficients
%$\{0.130602,0.496256,0.884392,-0.555356,0.69069\}$ to 
%network 1.
%
}
\label{fig:distinguishability-network}
\end{figure*}

We study the distinguishability of the interaction matrix by defining the \emph{interconnection vector} of node $i$ as ${\bm a}_i=(a_{i1}, \cdots, a_{in})^\T
\in \mathbb R^n$,  which is just the transpose of $A$'s $i$-th row. %, characterizing how node $i$ is affected by %interacts with 
%connects to
%the rest of the nodes. 
%
We also define the \emph{regressor vector} ${\bm f}_i (\bm x) =
(f_{i1}(x_i, x_1), \cdots,$ $ f_{in}(x_i, x_n) )^\T$ of
node $i$, characterizing the  coupling functions associated to node $i$.   
Then \eqref{system} can be rewritten as  
\begin{equation}
\label{system3}
\dot x_i(t) = {\bm f}_i^\T\big( \bm x(t) \big) {\bm a}_i + u_i(t), \quad
i=1,\cdots,n. 
\end{equation}
with $\bm x = (x_1, \cdots, x_n)^\T \in \mathbb R^n$ the state vector. Using this notation, the distinguishability of  $\mathcal P(A)$ is equivalent to the distinguishability of $\mathcal P(\bm a_i)$ for $i =1,\cdots,n$.  

%\textcolor{red}{Should I write here that uncertainity in $\bm f_i$ is handled using a group and family?}
%When we have uncertainty of the true system dynamics, we cannot 

In many cases, due to our lack of knowledge of the exact coupling functions, we may not know the true regressor $\bm f_i$ but only a \emph{family} of regressors $\{\bar {\bm f}_i \}$ to which it belongs.
%
%For example, the family of regressors that describe all systems with pairwise interactions. 
%
Members of the family can be considered as \emph{deformations} $\bar {\bm f}_i(\bm x)=\bm g_i (\bm f_i (\bm x))$ of the true regressor $\bm f_i(\bm x)$  obtained by applying some transformation $\bm g_i: \mathbb R^n \rightarrow \mathbb R^n$.
%of the true system dynamics $\bm f_i(\bm x)$ obtained using some transformation $G_i^*: \mathbb R^n \rightarrow \mathbb R^n$. 
%
This family can be characterized by a set  $ {\sf G}_i^*$ of admissible transformations, specified as follows: (i) this set is a \emph{group} \cite{fraieigh1994first},  and (ii)  any $\bm g_i \in {\sf G}_i^*$   is a continuous function that preserves pairwise interactions.
%
%The dual $G_i^*$ of  $G_i$ is defined as any  transformation that satisfies $[G_i^*(\bm f)]^\T \bm v = \bm f^\T G_i(\bm v)$ for all $\bm f, \bm v \in \mathbb R^n$.
%
%With \eqref{system3}, this family of regressors describes the family of dynamic systems with pairwise interactions.
%
Consider also the group ${\sf G}_{i, {\sf lin}}^*$ of linear transformations that preserve pairwise interactions.
 These linear transformations can be associated with nonsingular matrices $G_i^\T \in \mathbb R^{n \times n}$ with nonzero entries only in its diagonal and $i$-th column (see Fig.\ref{fig:distinguishability-network}a and SI-2).
Let  ${\sf G}_{i,{\sf lin}}$ denote the transpose of ${\sf G}_{i, {\sf lin}}^*$, i.e. $G_i \in {\sf G}_{i,{\sf lin}}$ if and only if $G_i^\T \in {\sf G}_{i,{\sf lin}}^*$.
Hereafter we use the following observation: since    ${\sf G}_{i, {\sf lin}}^* \subset {\sf G}_i^*$, a \emph{necessary} condition to reconstruct a property when $\bm g_i \in  {\sf G}_i^*$ is that it can be reconstructed when $\bm g_i \in  {\sf G}_{i, {\sf lin}}^*$.
Consequently, in order to characterize the fundamental limitations of network reconstruction, we  can focus on linear transformations only. We will show that linear transformations  are enough to produce severe limitations in the properties that can be reconstructed.
Using the notion of structural stability,  we later discuss the effects of deformations that do not belong to ${\sf G}_i^*$.

\subsection{Indistinguishable interconnection vectors}

Two candidate interconnection vectors $\bm v_1, \bm v_2 \in \mathbb R^n$ will be indistinguishable if they produce the same right-hand side in  Eq.  \eqref{system3} for some regressor in the family $\{\bar {\bm f}_i \}$.
%
%In particular, $\bm v_1$ is indistinguishable from $\bm v_2$ if
This is equivalent to the condition
\begin{equation}
\label{indist-general}
\bm f_i^\T( \bm x(t)) \bm v_1 = \bm f_i^\T( \bm x(t)) G_i \bm v_2, \quad \forall t \in [t_0, t_1],
\end{equation}
for  some  matrix $G_i  \in {\sf G}_{i,\sf lin}$, where $\bm x(t)$ is the measured node trajectories.
Multiplying this equation 
by ${\bm f}_i( \bm x(t))$ from the left and integrating
over the time interval $[t_0, t_1]$ we obtain 
\begin{equation}
\label{indistinguishable-vectors}
M_i(t_0,t_1)  ({\bm v}_1 - G_i {\bm v}_2) = {\bm 0}, %\quad M_i(t_1)= \int_{0}^{t_1} \Gamma_i(t) \Gamma_i^T(t) dt.
\end{equation}
where 
%\be 
$M_i(t_0,t_1) = \int_{t_0}^{t_1} {\bm f}_i(\bm x(t)) {\bm
  f}_i^\T( \bm x(t)) \, \ud t
$ 
is a constant $n \times n$ matrix. 
It is obvious that \eqref{indist-general} implies \eqref{indistinguishable-vectors}, but the converse implication is not so obvious  (Proposition 1 of SI-3). Indeed,  it constitutes the main obstacle  to extend our analysis to more general uncertainty of the coupling functions. 
Hereafter we write $M_i$ instead of $M_i(t_0, t_1)$,
unless the specific time interval is important for the discussion.   
From \eqref{indistinguishable-vectors},
the set of all pairs of indistinguishable interconnection vectors for node $i$ is given by
\begin{equation}
%\begin{split}
\Omega_i = \left \{({\bm v}_1, {\bm v}_2) \in \mathbb R^n \times
  \mathbb R^n \Big | \exists  G_i \in {\sf G}_{i, \sf lin} \ \mbox{such that}\ ({\bm v}_1 - G_i {\bm v}_2) \in \ker M_i \right \}. \\
 %=&  \left \{v_1, v_2 \in \mathbb R^N / \ker M_i(t_1) \right \}
%\end{split}
\end{equation}

\begin{figure*}[!t]
\centering
\includegraphics[width=7in]{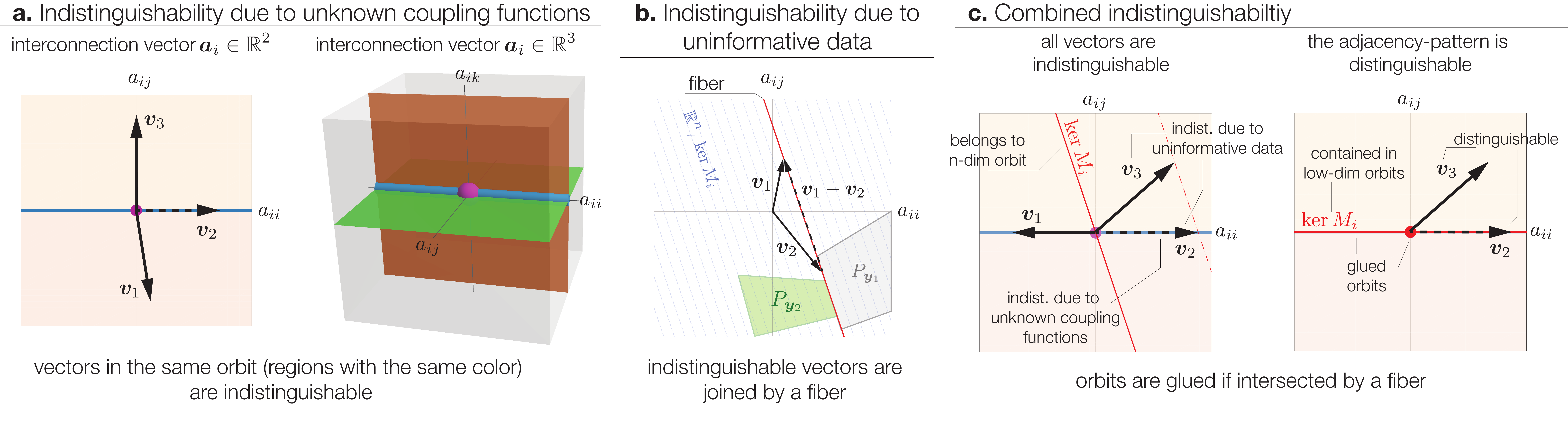}
\vspace*{-0.3cm}
\caption{\small 
% {\bf Indistinguishable networks prevent network
%     reconstruction.}
{\bf Indistinguishability of interconnection vectors.} 
{\bf a.} 
Indistinguishable vectors due to unknown coupling functions can be transformed into each other using some transformation $G_i \in {\sf G}_{i, \sf lin}$.
Sets of those indistinguishable vectors are the partition of $\mathbb R^n$  by the orbits $\mathcal O_i$ of ${\sf G}_{i, \sf lin}$, here shown in different colors for $n=2$ and $n=3$.
Purple regions should be interpreted as points, and blue regions as lines. The grey region is another orbit.
We can distinguish an interconnection vector in the blue orbit (e.g., $\bm v_2$  with component $a_{ij}=0$)  from an interconnection vector in the orange orbit (e.g., $\bm v_1$ or $\bm v_3$ with component $a_{ij}\neq 0$), illustrating that we can distinguish the adjacency of the interconnection vector (i.e., wether $a_{ij}$ is zero or not for $j \neq i$). 
Nevertheless, since $\bm v_1$ and $\bm v_3$ belong to the same orbit and hence are indistinguishable, but they have different degree sequences and sign or connectivity patterns, 
% $\bm v_2$ and $\bm v_3$ have different degree sequences and sign/connectivity patterns but are indistinguishable, 
these properties cannot be reconstructed.  
{\bf b.}
Due to uninformative measured temporal data, the interconnection vector $\bm v_1$ is indistinguishable from $\bm v_2$ because $\bm v_1 - \bm v_2 \in \ker M_i$, that is, both vectors are joined by a fiber (shown in red).
 Note also that we can separate the sets   $P_{{\bm y}_1}$ and $P_{{\bm y}_2}$ with the particular orientation of the fibers.
 However, since there is no gap between these sets, any change in the orientation of the fibers (regardless of how small it is) will produce indistinguishable interconnection vectors that belong to different sets.
 This illustrates that the PE condition remains generically necessary if there is no gap between the sets in $\mathcal C_{\mathcal P} = \{ \mathcal P^{-1} (\bm y) \subseteq \mathbb V | \bm y \in \mathbb Y \}$.
{\bf c.}
Indistinguishable vectors in network reconstruction appear by combining both kinds of indistinguishable vectors,  gluing together orbits of ${\sf G}_{i, \sf lin}$ when they are intersected by a fiber of $\mathbb R^n / \ker M_i$.
In the left panel, since $\ker M_i$ is not contained in low-dimensional orbits, all orbits are  are glued ${\mathcal O}_i^{\ker M_i} = \mathbb R^2$ and  all vectors become indistinguishable (e.g., $\bm v_1$ is indistinguishable from $\bm v_3$).
In the right panel,  $\ker M_i$ is horizontally oriented and hence contained in low-dimensional orbits. We can then 
distinguish between $\bm v_2$ and $\bm v_3$ and hence reconstruct the adjacency pattern of the interaction matrix. 
% distinguish the adjacency-pattern of the interconnection vector and thus distinguish between $\bm v_2$ and $\bm v_3$.
 %because $\bm v_1$ is indistinguishable from $\bm v_2$ (due to unknown dynamics), %because they belong to the same orbit;  
%and $\bm v_2$ is indistinguishable from $\bm v_3$ (due to uninformative data). %Consequently, $\bm v_1$ is  also indistinguishable  from $\bm v_3$.
%
}
\label{fig:distinguishability-sources}
\end{figure*}

The above equation shows two sources of indistinguishability, rendering two classes of fundamental limitations of NR. 
First, unknown coupling functions causes two vectors to be indistinguishable if they can be transformed to each other via some $G_i \in {\sf G}_{i,\sf lin}$.
This set of indistinguishable vectors $\{(\bm v_1,\bm v_2)|\bm v_1=G_i \bm v_2, G_i \in G_{i,\sf lin} \}$ is then the partition ${\mathcal O }_i$ of $\mathbb R^n$ by the \emph{orbits}  of the group ${\sf G}_{i,\sf lin}$ \cite{fraieigh1994first}, Fig. \ref{fig:distinguishability-sources}a and SI-2.
An orbit is called low-dimensional if its dimension is $<n$ (purple, blue, green and brown orbits in Fig.2a).
The orbits in Fig.\ref{fig:distinguishability-sources}a show that unknown coupling functions allow us to distinguish only if $\dot x_i$ depends on $x_j$ for $j \neq i$ (i.e. the adjacency pattern of the interconnection vector), see Proposition 2a and Example 1 in SI-5. 
This is a consequence of the invariance of the adjacency pattern to the transformations in ${\sf G}_{i}^*$ (i.e. prior knowledge of the coupling functions).
Other  properties like edge-weights, connectivity patterns  or degree sequence are indistinguishable and cannot be reconstructed (Proposition 2b in SI-5). %See Proposition 1 on SI-1.3 for the formal proof.
Second,  even if $\bm f_i(\bm x)$ is exactly known, indistinguishability can still emerge due to 
%indistinguishability \emph{intrinsic} to the system dynamics $\bm f_i(\bm x)$ and the data $\bm x(t)$ 
 uninformative data (Fig. 1b),  making  $\bm v_1$ indistinguishable from $\bm v_2$ if $\bm v_1 - \bm v_2   \in \ker M_i$ and $\ker M_i$ is nontrivial (i.e. contains a linear subspace different from $\bm 0$).
In other words,   the endpoints of $\bm v_1$ and $\bm v_2$ can be connected by an hyperplane parallel to $\ker M_i$, Fig.\ref{fig:distinguishability-sources}b. 
Note that hyperplanes parallel to $\ker M_i$ are often called \emph{fibers} of the quotient space $\mathbb R^n / \ker M_i$.

Combining these two sources of indistinguishability,  ${\bm v}_1$ is indistinguishable from ${\bm v}_2$ iff it is possible to transform $\bm v_2$ using an element of ${\sf G}_{i,\sf lin}$ in a way that the line (or more generally, hyperplane) passing trough $\bm v_1$ and ${\sf G}_{i,\sf lin} \bm v_2$ is a fiber. 
Consequently,  orbits of ${\sf G}_{i,\sf lin}$ intersected by a fiber of  $\mathbb R^n / \ker M_i$ become  indistinguishable and we can `glue' them together to form a partition ${\mathcal O}_{i}^{\ker M_i}$ of $\mathbb R^n$ into sets of indistinguishable interconnection vectors, Fig. \ref{fig:distinguishability-sources}c.
If $\ker M_i$ is not contained in  low-dimensional orbits, then ${\mathcal O}_{i}^{\ker M_i} = \mathbb R^n$ and all vectors are indistinguishable.
 If, however, $\ker M_i$ is contained in low-dimensional orbits then we can reconstruct the adjacency pattern of the interaction matrix (right panel of Fig. 2c).
Note that the partition of indistinguishable interconnection vectors due to the nonlinear deformations cannot be finer than ${\mathcal O}_{i}^{\ker M_i}$ obtained via linear deformations.

Note also that the matrix $M_i$  in \eqref{indistinguishable-vectors} is typically unknown because the true regressor $\bm f_i$ is unknown.  Certainly, choosing any regressor $\bar {\bm f}_i= G_i^\T \bm f_i$, $G_i \in {\sf G}_{i,\sf lin}$, we can only compute 
 $$\bar  M_i(t_0, t_1) = \int_{t_0}^{t_1}  \bar {\bm f}_i( \bm x(t) ) \bar {\bm
  f}_i^\T( \bm x(t) ) \, \ud t = G_i^\T M_i(t_0, t_1) G_i.$$
 Therefore, we have only access to properties of $M_i$ that remain \emph{invariant} under $G_i^\T M_i G_i$ for any  $G_i \in {\sf G}_{i,\sf lin}$.
To find those invariant properties, note that if $\bar {\bm v} \in \ker \bar M_i$ then $\bm v = G_i \bar{ \bm v} \in \ker M_i$, since $\bm 0 = \bar M_i \bar {\bm v} = G_i^\T M_i G_i \bar {\bm v}$ and $G_i^\T$ has full rank. Thus, ${\sf G_{i,\sf lin}}$ transforms  $\ker \bar M_i$ into $\ker M_i$ (and vice-versa, because it is a group), and
  we can only know the orbit ${\sf G}_{i,\sf lin}(\ker M_i)$ corresponding to this subspace. 
  %ell only if the subspace $\ker M_i$ belongs to some orbit of ${\sf G}_i$ or not. 
%
   For example, the condition $\ker \bar M_i = \{ \bm 0 \}$ for some $G_i \in {\sf G}_{i,\sf lin}$ implies that  
\begin{equation}
\ker M_i =\{{\bm 0}\},  \label{PE}
\end{equation}
because ${\sf G}_{i, \sf lin}( \bm 0)= \bm 0$ (i.e., $\bm 0 = G_i \bm 0$ for any $G_i$). This shows   that we can tell if $M_i$ is nonsingular using any $\bar M_i$. 
 Equation \eqref{PE} is an important condition in system identification literature known as \emph{Persistent Excitation} (PE) and it is necessary and sufficient  to solve the classical PI problem \cite{narendra2012stable}.   
 With \eqref{PE} the data is informative enough in the sense it does not produce indistinguishability.
In general,  we can build the partition of indistinguishable vectors  using any $\bar M_i$, i.e., $\mathcal O_i^{\ker \bar M_i} = \mathcal O_i^{\ker M_i}$ (see Lemma 1 of SI-6 for the proof).

%%%%%%%%%%%%
%%%%
%%%%
%%%%
%%%%%%%%%%%%

\subsection{Necessary condition to distinguish a property} Let %$\mathcal{P} : {\mathbb V} \subseteq \mathbb R^n \rightarrow \mathbb Y$
$\mathcal{P} :  \mathbb R^n \rightarrow \mathbb Y$ be the
property of the interconnection vector we want to reconstruct, where
 $\mathbb Y$ is its image. For example, $\mathbb Y
= \{-1,0,1\}^n$ if $\mathcal P$ is the sign pattern, or $\mathbb Y = \{0,1\}^n$ if
$\mathcal P$ is the adjacency or connectivity pattern.   
%
%The domain $\mathbb V$ becomes smaller than $\mathbb R^n$ when we use prior information to
%discard some interconnection vectors. % (see Sec \ref{sub-priorinfo}). 
%
%
%
The property ${\mathcal P}$ can be reconstructed only if any two interconnection vectors
$\bm v_1, \bm v_2 \in \mathbb  R^n$ that have different properties $\bm
y_1 = {\mathcal P}(\bm v_1) \neq {\mathcal P} (\bm v_2) = \bm y_2$ are
distinguishable, i.e., belong to different orbits of ${\mathcal O}_i^{\ker M_i}$.   
Let $P_{\bm y} = \mathcal{P}^{-1}(\bm y) = \{ \bm v \in \mathbb R^n | \mathcal P (\bm
v) = \bm y \}$. % be the
%set of all interconnection
%vectors $\bm v \in \mathbb V$ that satisfy ${\mathcal P}(\bm v) = \bm
%y$. % , that is  $P_{\bm y} = {\mathcal P}^{-1}(\bm y)$ for all $\bm y \in
% \mathbb Y$.   
Then % all different values $\bm y \in \mathbb Y$  for ${\mathcal
  % P}(\bm v)$ can be distinguished, or equivalently, 
 ${\mathcal P}(\bm
a_i)$ can be reconstructed only if all two sets in the collection
$\mathcal C_{\mathcal P}=\{P_{\bm y} \subseteq \mathbb R^n | \bm y \in \mathbb
Y\}$  belong to different orbits ${\mathcal O}_i^{\ker M_i}$. %
When the deformations are a-priori known to be linear, this condition is also sufficient.

%%%%%%%%%%%%
%%%%
%%%%
%%%%
%%%%%%%%%%%%

\subsection{The role of our knowledge of the coupling functions}  We could shrink or enlarge the group of transformations ${\sf G}_{i, \sf lin}$ according to our uncertainty of the coupling functions.  
For example, it will collapse to the single element ${\sf G}_{i, \sf lin} = \{ I_{n \times n} \}$ if we know the coupling functions exactly, or will increase to ${\sf G}_{i, \sf lin} = \{$nonsingular  $\mathbb R^{n \times n}$ matrices$\}$ if we do not have any knowledge of them.
We emphasize that if  ${\sf G}_{i, \sf lin}$ is enlarged (e.g., by including nonlinear transformations or more general interactions between nodes), existing  orbits may merge but new orbits cannot appear because the original linear transformations preserving pairwise interaction remain in the group.
%
%For example, we .%,  and has a single orbit $\mathcal O_i=\mathbb R^n$.

Since our previous analysis only depends on the group property of the transformations, it can be straightforwardly extended  to  any linear group ${\sf G}_i$. It is just necessary to find  its orbits $\mathcal O_i$ and build the corresponding ${\mathcal O}_i^{\ker M_i}$. 
From this observation, in order to reconstruct some property of the interaction matrix,  it is necessary that (i) our uncertainty about the coupling functions is  small enough (i.e., any two sets in $\mathcal C_{\mathcal P}$ belong to different orbits of ${\sf G}_i$),  and (ii) the measured temporal data is informative enough (i.e., hyperplanes parallel to ${\sf G}_i(\ker M_i)$  do not glue orbits together).
For example, in order to reconstruct the edge-weights it is necessary to know the coupling functions exactly (${\sf G}_i=\{I\}$), because only then any two vectors belong to different orbits. 

%%%%%%%%%%%%
%%%%
%%%%
%%%%
%%%%%%%%%%%%

\subsection{Specifying the coupling functions} It is possible to reduce   ${\sf G}_i$ to $\{I\}$ when the system we aim  to model indicates the appropriate coupling functions to use.
For example, the generalized Lotka-Volterra (GLV) model can provide a good starting point for ecological systems  \cite{barzel2013universality}.
Linear coupling functions are appropriate if the system remains close to an operating point (e.g., a steady-state).
Candidate coupling functions for the model can  also be computationally searched or improved using symbolic regression \cite{bongard2007automated}.
 In these cases indistinguishability  emerges only  from uninformative data: $\bm v_1$ is indistinguishable from $\bm v_2$ iff $\bm v_1 - \bm v_2 \in \ker M_i$. 
Consequently, % all different values $\bm y \in \mathbb Y$  for ${\mathcal
  % P}(\bm v)$ can be distinguished, or equivalently, 
a property ${\mathcal P}(\bm
a_i)$ can be reconstructed iff all two sets in the collection
$\mathcal C_{\mathcal P}=\{P_{\bm y} \subseteq \mathbb R^n | \bm y \in \mathbb
Y\}$ can be \emph{separated} by a fiber, Fig. S1.
A fiber is an hyperplane and thus partitions $\mathbb R^n$ in two regions; we say it separates $P_{{\bm y}_1}$ from $P_{{\bm y}_2}$ if 
$P_{{\bm y}_1}$ belong to one region and $P_{{\bm y}_2}$ belongs to the other region or the fiber, Fig. 2b.

By specifying the coupling functions we can reconstruct more information such as the interaction matrix itself (i.e., edge-weights). Setting $\mathcal P ={\sf Identity}$ we obtain $\mathcal C_{\mathcal P} = \mathbb R^n$, showing that the necessary and sufficient condition to
reconstruct $A$ %the edge-weight 
is to distinguish between any two different  interconnection vectors 
% points 
in $\mathbb R^n$. 
This is possible iff the PE condition \eqref{PE} holds, a classical result from system identification theory \cite{narendra2012stable}.
Without PE it is still possible to distinguish, for example,  the adjacency-pattern of the interconnection vector when $\ker M_i$ is exactly `horizontally' oriented. In fact, 
from the right panel of Fig. 2c, we can separate the sets $P_{\bm y}$ of vectors with different adjacency-patterns (orange and red regions) using the same red region as separating fiber.
However, this situation is pathological in the sense that an infinitesimal change in the fiber's orientation will eliminate the distinguishability.
Note also that other properties like sign-pattern, connectivity pattern or degree sequence are indistinguishable.

%%%%%%%%%%%%%%%
%%%%%
%%%%%
%%%%%
%%%%%%%%%%%%%%%

\subsection{Persistent excitation is generically necessary} Any mathematical model  only approximates the dynamic behavior of a real system.
Therefore, we can only expect  that the ``true'' coupling functions are sufficiently close (but not exactly equal) to some deformation $\bar {\bm f}_i(\bm x) = G_i^\T \bm f_i(\bm x)$, $G_i \in {\sf G}_{i, \sf lin}$.
Considering this, it is important to understand if the distinguishability conditions derived earlier remain true under arbitrary but sufficiently small deformations of the coupling functions, a notion known as  \emph{structural stability} \cite{thom1989structural,golubitsky1978introduction}.
Otherwise, these conditions represent  \emph{non-generic} cases that cannot  appear in practice because they vanish under infinitesimal deformations.

We proved that the PE condition \eqref{PE} is  structurally stable (Theorem 1, SI-7). 
However,  when $\ker M_i$ is non-trivial, the condition that it belongs to low-dimensional orbits is structurally unstable (Theorem 2, SI-7). 
%
%This implies the following. 
%
To understand the implications of these results,  let's consider  an arbitrary deformation $\hat {\bm f}_i(\bm x)$ with `size' $\delta >0$, i.e., $\| \hat {\bm f}_i(\bm x(t)) - \bar{\bm f}_i( \bm x(t)) \| \leq \delta, \forall t \in [t_0, t_1]$.
%
%This deformation may be interpreted as the `true' dynamics of the system.
%
The PE condition is structurally stable because there exists $\delta >0$ sufficiently small such that if $\bar{\bm f}_i( \bm x(t))$ has PE then any $\hat {\bm f}_i(\bm x (t))$ also has PE, Fig.3a.
Indeed, regardless of the size of the deformation, almost any analytic deformation of the regressor will also have PE (Theorem 3, SI-8).
In practice, these two results imply that we can check if given temporal data satisfies the PE condition without knowing the coupling functions exactly.
In contrast, when $\ker M_i$ is non-trivial (i.e., contains a linear subspace of $\mathbb R^n$ different from $\bm 0$) and belongs to low-dimensional orbits, then for any $\delta >0$ there is a deformation $\hat {\bm f}_i(\bm x)$ ---a rotation, indeed--- such that $\ker \hat M_i$ belongs to the $n$-dimensional orbit, Fig.3b. Here $\hat M_i = \int_{t_0}^{t_1} \hat {\bm f}_i(\bm x (t)) \hat {\bm f}_i^\T(\bm x (t)) \ud t$.

The analysis above shows that only two \emph{generic} cases exist:
%\begin{itemize}
(i) $\ker M_i = \{\bm 0\}$ and  indistinguishable vectors emerge only due to uncertain coupling functions $\mathcal O_i^{\ker M_i} = \mathcal O_i$; and %Therefore, we can only reconstruct properties that are distingsuihable  
(ii) $\ker M_i$ is not trivial and  is contained in the $n$-dimensional orbit,  so all interconnection vectors become indistinguishable $\mathcal O_i^{\ker M_i} = \mathbb R^n$.

\begin{figure}[!t]
\centering
\includegraphics[width=3in]{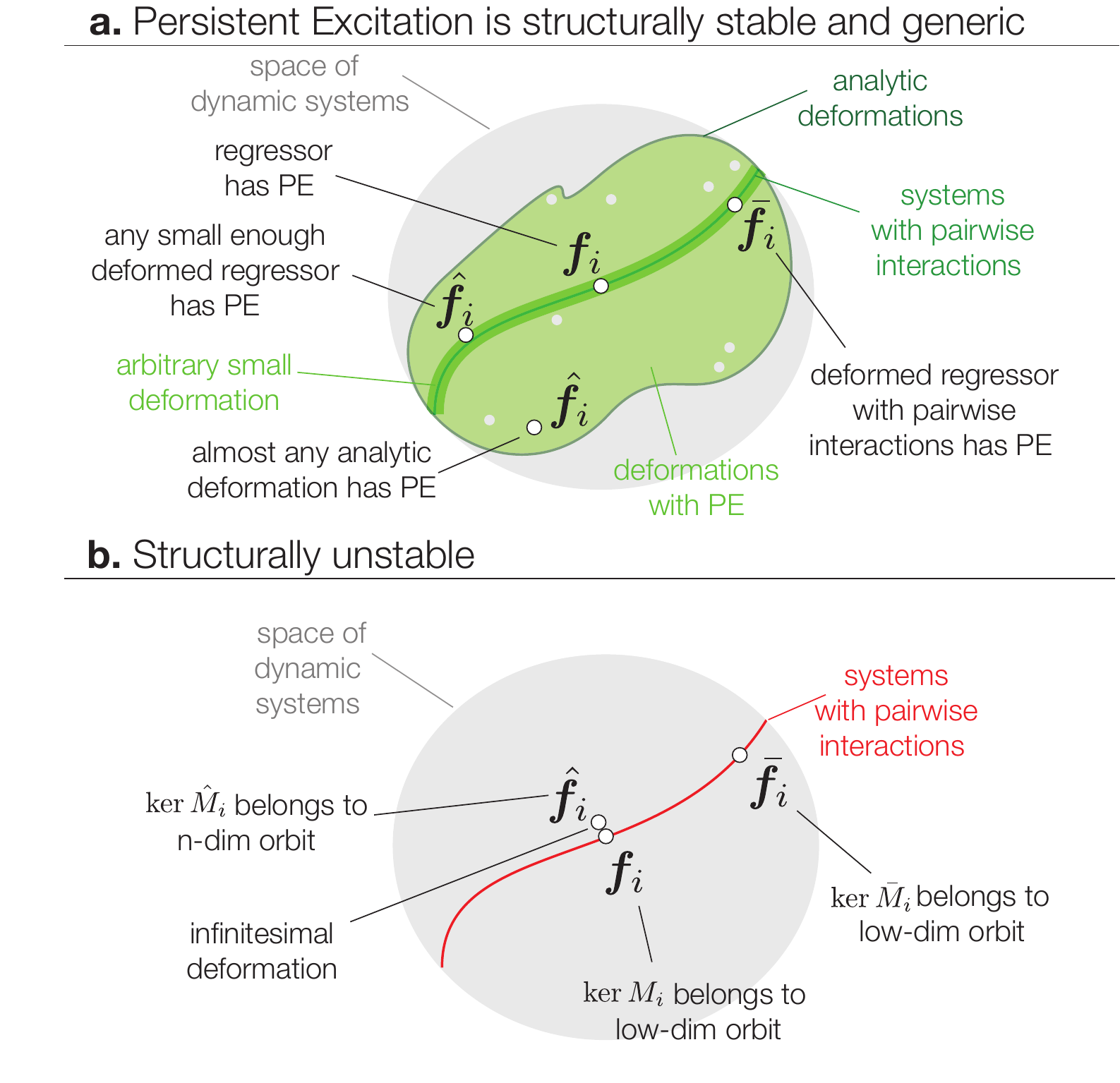}
\vspace*{-0.0cm}
\caption{\small 
{\bf Schematic illustration of structural stability.} 
{\bf a.} The Persistent Excitation condition \eqref{PE} is structurally stable because once  some regressor $\bar {\bm f}_i$ has PE, any small enough deformation $\hat {\bm f}_i$ of it also has PE.
{\bf b.} When $\ker M_i$ is nontrivial, the condition that it belongs to low-dimensional orbits is structurally  unstable because there always exists a infinitesimal deformation $\hat {\bm f}_i$ such that $\ker \hat M_i$ belongs to the $n$-dimensional orbit.%
}
\label{fig:structural-stable}
\end{figure}

Consequently, in a generic case,  the PE condition \eqref{PE} is necessary   in order to reconstruct any property. 
Even if the coupling functions are exactly known,  without PE we cannot  \emph{generically} reconstruct    the sign/connectivity/adjacency patterns or degree sequence. The reason is simple: for all these properties there is no gap between the sets $\mathcal C_{\mathcal P}$. 
For example, for $\varepsilon \approx 0$,  the vectors $\bm v_1=(\varepsilon, 0, \cdots, 0)^\T$ and $\bm v_2 = \bm 0$  are infinitesimally close in $\mathbb R^n$ but have different connectivity or degree sequence.
Therefore, even when the sets $P_{\bm y}$ can be separated by a fiber with a particular orientation (e.g,  $P_{\bm y_1}$ and $P_{\bm y_2}$ shown  in  Fig.\ref{fig:distinguishability-sources}b), 
an infinitesimal deformation in the coupling functions changes this orientation producing indistinguishable interconnection vectors with different properties.  The question is how to create these gaps and solve NR problems without PE.

In the following,  we show that knowing prior information about the interaction
matrix $A$  shrinks the domain of a property $\mathcal
P$, create gaps between the sets $P_{\bm y}$ in $\mathcal C_{\mathcal P}$ and hence 
relax the PE condition. 

%%%%%%%%%%
%%%%%%%%%%
%%%%%%%%%%%%%%%%%%%%
%%%%%%%%%%

\subsection{Prior knowledge of the interaction matrix relaxes the PE condition}
For clarity, in this section we assume that the coupling functions are exactly known. 
The simplest 
prior information of $A$ is %by constraints 
that the interconnection vectors satisfy:%
% Prior knowledge of the network can be represented as sets $Q_i
% \subseteq \mathbb R^n, i=1,\cdots, n$, constraining the possible
% edge-weights of the interconnection vector 
\begin{equation}
\label{constraint}
{\bm a}_{i} \in \mathbb V, \quad i=1,\cdots, n,
\end{equation}
where 
 $\mathbb V \subseteq \mathbb R^n$ is a known set. 
 Prior information shrinks the domain of the property $\mathcal P$  from $\mathbb R^n$ to $\mathbb V$, i.e., $\mathcal P : \mathbb V \subseteq \mathbb R^n \rightarrow \mathbb Y$.
Two typical cases are: (i) $a_{ij}$ takes a finite number of
values (e.g., binary signed interactions) 
 and $\mathbb V = \cup_{\bm y} P_{\bm y}$ is a discrete set since each $P_{\bm y}$ is a point, Fig.4a; and (ii)  $a_{ij}$ are bounded as
\begin{equation}
 \label{bounds}
 a_{ij} \in [-a_{\max}, -a_{\min}] \cup [-\epsilon, \epsilon] \cup
 [a_{\min},a_{\max}] 
 \end{equation}
for some known constants $0 \leq
\epsilon<a_{\min}<a_{\max}$.
In this case  $\mathbb V = \cup_{\bm y} P_{\bm y}$,  where
$P_{\bm 0}$ is an $\epsilon$
neighborhood of zero (which can be associated to `zero' sign-pattern), and each of the $3^n-1$ remaining sets lies in a
different orthant $\mathbb R^n$ (and thus be associated to distinct sign patterns),
see Fig.4b.   
A similar analysis can be applied in the case when `network sparsity' is the prior information, SI-9.

In case (i), $A$ itself can be reconstructed without PE if we can separate each point composing $\mathbb V$ with a fiber.
If $\dim (\ker M_i) <n$, this is generically possible because an infinitesimal deformation will change any `pathological' orientation that contains two points.
%
%we can reconstruct $A$ itself without PE \textcolor{blue}{when}  it
%is possible to separate each point composing $\mathbb V$ with a fiber parallel
%to $\ker M_i$. Indeed,  under prior (\ref{constraint}) it is a generic condition
%that the link weights can be reconstructed, in the sense that the
%probability that a  hyperplane with random orientation contains two given
%points is zero. 
%
In case (ii), the sign or connectivity pattern can be reconstructed without PE if there is a gap between the sets $P_{\bm y}$ such that a  fiber can separate them, Fig.4b. 
The condition that a fiber fits in a gap is structurally stable.
If this gap increases ($a_{\min} - \epsilon$ increases and  $a_{\max} - a_{\min}$ decreases), it becomes even easier  for the fibers to fit. 
However, the interaction matrix $A$ itself cannot be
reconstructed because it is impossible to separate two points inside
one $P_{\bm y}$.
%Indeed, the link weights  cannot  be
%estimated with an error lower than 
%$\max\{a_{\max}-a_{\min}, \epsilon\}$. 

\begin{figure}[!t]
\centering
\includegraphics[width=3.3in]{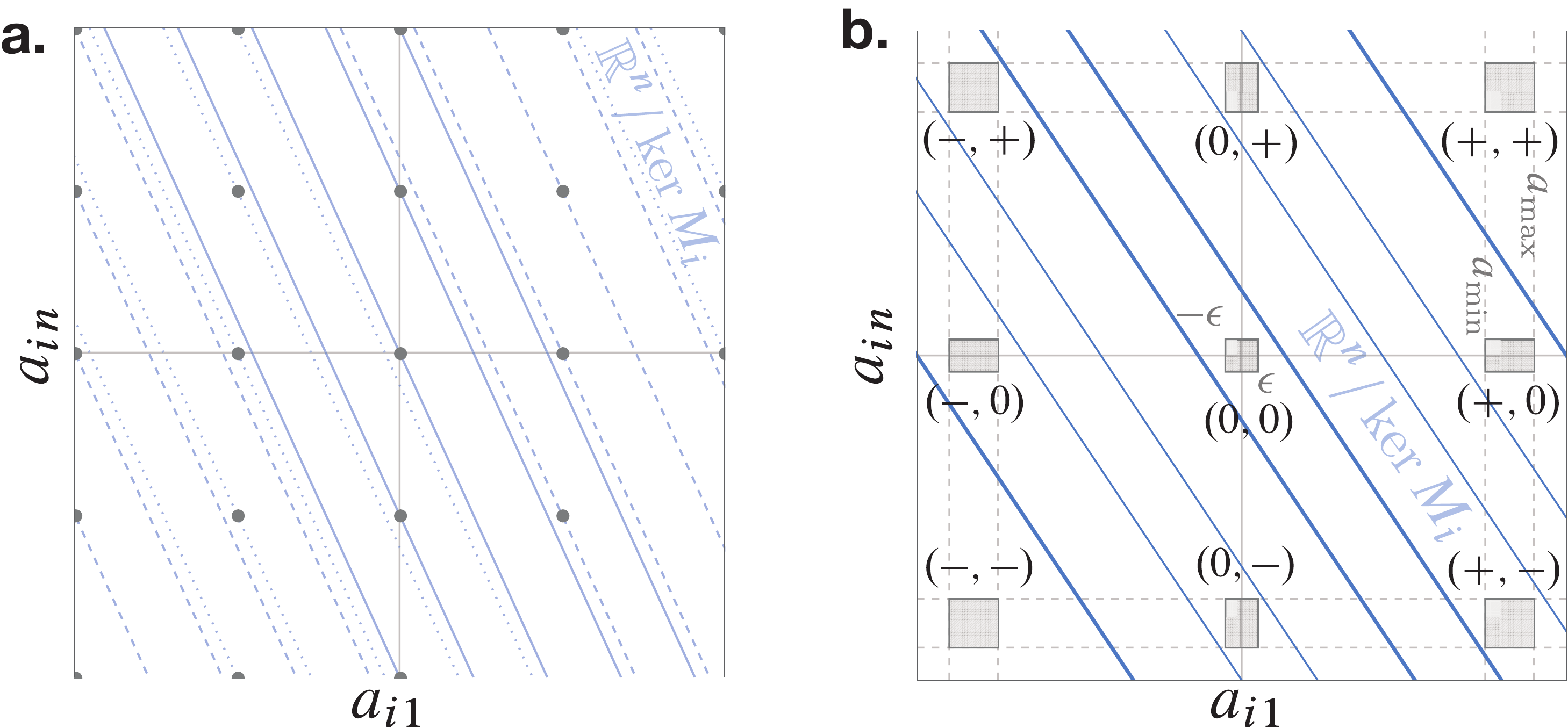}
\vspace*{0.1cm}
\caption{ \small {\bf Prior information of the interaction matrix relaxes the PE condition.} 
%{\bf a.} The
%  two interconnections $\bm v_1, \bm v_2 \in \mathbb R^n$ are
%  indistinguishable since $(\bm v_1 - \bm v_2) \in \ker M_i$. In
%  other words, they are indistinguishable since the line connecting
%  them (shown in red) is parallel to $\ker M_i$, and it is a fiber of
%  $\mathbb R^n / \ker M_i$. 
%% Therefore, the sets $P_1$ and $P_3$ are indistinguishable. 
%    %
%    It is impossible to find a hyperplane separating  $P_+ = \{\bm v \in \mathbb R^n | v_1 >0 \mbox{ and } v_n >0\}$ from $P_0 =  \{\bm v \in \mathbb R^n | v_1 >0 \mbox{ and } v_n = 0\}$, or $P_+$ from  $P_- = \{\bm v \in \mathbb R^n | v_1 >0 \mbox{ and } v_n <0\}$, or $P_+ \cup P_-$ from $P_0$. 
%        %
%       Therefore, it is necessary that $\ker M_i$ is the trivial
%       vector $\bm 0_{n}$, which is the PE condition, in order to
%       distinguish either the sign pattern or connectivity pattern of the interconnection vector.
        %
 {\bf a.} When the edge-weights $a_{ij}$'s take a finite-number of
 values, the set $\mathbb V$ is discrete (shown in grey). Then distinguishability of
 the edge-weights is generic, because an infinitesimal deformation will change any fiber that contains two elements of $\mathbb V$ (grey points). 
{\bf b.} Example for the sets $P_{\bm y}$ (shown in grey) in the case
 of known bounds of the edge-weight \eqref{bounds}. 
 Though the
 edge-weights cannot be distinguished, the sign-pattern (and
 connectivity) can still be distinguished since there exist hyperplanes
 parallel to $\ker M_i$ (shown in blue) separating every $P_{\bm y}$. This condition is structurally stable. }
\label{fig:dist-diagram}
\end{figure}

%%%%%%%%%%%%
%%%%
%%%%%%%%
%%%%
%%%%%%%%%%%%

\subsection{Example} 
We illustrate our results in a basic problem of network reconstruction using steady-state data. Consider two species $(x_1,x_2)$ interacting in a food web and suppose we  measure their steady-state abundances 
$\bm x(t)=\mbox{const},  \forall t \in [t_0, t_1]$.
The goal is to reconstruct the sign-pattern of the interaction matrix characterizing who eats whom.
Since the data is constant, any regressor $\bar{\bm f}_i(\bm x(t))$ is also constant and all $\bar M_i$'s have rank 1 at most. Thus, the PE condition \eqref{PE} cannot be satisfied.
Consequently, our analysis shows that without better specifying the coupling functions of the model,  all interconnection vectors are generically indistinguishable and it is impossible to reconstruct any property of the interaction matrix. 
%
%This happens because, generically, the line $\ker \bar M_i$ will not align with the orbit $\spa(1,0)^\T$ shown in blue in Fig.\ref{fig:distinguishability-sources}a.%, see also SI-??.

To circumvent this problem we specify the coupling functions using the GLV model $
\dot x_i = r_i x_i + \sum_{j=1}^2 a_{ij} x_i x_j$. We assume that the growth rates $r_i$ are known. 
This uncontrolled model can be rewritten as in \eqref{system} using  $f_{ij} (x_i, x_j) = x_i x_j$ and $u_i(t) = r_i x_i(t)$. 
Note that $M_i(t_0, t_1)=x_i^2 \cdot (t_1 - t_0) \cdot \bm x \bm x^\T$ has rank $1$ at most, and it is still generically impossible to reconstruct exactly the interaction matrix $A=(a_{ij})$ or any other property of it. 
This coincides with the fact that one steady-state experiment is generically not enough for parameter identification \cite{sontag2002differential}.
Yet, assuming known  bounds of the interactions \eqref{bounds}, we can reconstruct exactly the sign-pattern. 
For this, it is necessary and sufficient to separate the $3^2$ sets in $\mathcal C_{\mathcal P}$ by lines parallel to $\ker M_i$. 
%
%Since the sets $P_{k}$ are polyhedral, it is numerically efficient to decide if a fiber exists or not (SI-\ref{SI-separating-hyperplanes}).
%
SI-10 presents a numerical example  when this is possible, and SI-4 shows an NR method based on our analysis.

\section{Discussion}

We now discuss  the implications of our results.
%\subsection{Advantages  of NR over PI} %NR differs from PI when we want to infer less information (e.g., adjacency instead of edge-weights), and intuition suggests this should translate into weaker necessary conditions to solve the problem.
%
 Regardless of the property of the interaction matrix $A$ we aim to reconstruct and even if we know the coupling functions exactly, we proved that PE  \eqref{PE} is generically necessary. 
 %Otherwise all vectors simply become indistinguishable under infinitesimally small uncertainty in the dynamics,  and no property of the interaction matrix can be reconstructed.   
This fundamental limitation implies that  reconstructing less information of the interaction matrix generically does not mean  we can solve an NR problem with less informative data. % (e.g., steady-state information only).
In particular, when only steady-state data from a single experiment is available,  our result implies that  generically no property of the interaction matrix can be reconstructed, not even mentioning the interaction matrix itself  \cite{sontag2002differential}.
%
%Indeed, we shown that only when prior information of the interaction matrix is available, NR problems can be solved without PE.
From a different angle, the PE condition also serves as guideline to design  experiments \cite{bandara2009optimal} that can provide sufficiently informative data.
For instance, simply changing the initial conditions of the two-species ecological network of our previous example can produce PE using the GLV model.
Available control inputs  and intrinsic noise on the dynamics are also useful for this \cite{shimkin1987persistency,Preciado:15b}.
Notice that in system identification literature PI is often performed in real time, so the PE condition should hold uniformly in the initial time \cite{ljung1998system,kailath2000linear,narendra2012stable}.

The advantage of using NR to reconstruct less information of the interaction matrix is that we can have more uncertainty about the system dynamics. 
For example, if we aim to reconstruct  the adjacency-pattern $K$ and the PE condition holds, we can consider   the set of all dynamic systems with pairwise coupling functions and  we need little knowledge of the true system dynamics.
We can check the PE condition even when the coupling functions are not exactly known  (SI-7 and SI-8).
Indeed, for linear deformations, we characterized an optimal tradeoff:  given a property of $A$ to reconstruct, the uncertainty on the coupling functions should be  small enough (orbits distinguish the property) and the measured data should be informative enough (to ensure PE in the family of regressors).
It remains open to understand how much indistinguishability is created by considering general nonlinear deformations.
%Yet, given $\mathcal P$,  it remains open to characterize the largest admissible uncertainty in the sense that the orbits of the group ${\sf G}_i$ distinguish any two sets in $\mathcal C_{\mathcal P}$.

%From the above discussion, it  also follows that any reconstructed sign-pattern or edge-weights should be  associated  to a family of dynamic models. Otherwise, the characterization is  incomplete because changing the family of models will  change the edge-weights and sign-pattern. 

Experimentally  measured data  usually has poor information content, in the sense that it typically cannot satisfy the PE condition. For example, current gene sequencing is frequently constrained to measure steady-state data only, which cannot satisfy the PE condition for \emph{any} regressor. 
In order to circumvent this fundamental limitation of NR, we have shown that prior knowledge of the interaction matrix can relax the PE condition allowing us to solve the NR problem.
%

%\subsection{Concluding remarks}

We notice that a different class of fundamental limitation in NR has been discussed in literature: solving an NR problem is impossible without measuring all time-varying nodes in the network \cite{gonccalves2008necessary}. %In other words, if there are some hidden nodes in the network that are time-varying, then NR is impossible. 
If the state variables of unmeasured nodes are constant, then NR is actually possible (SI-11).
Previous works considered  the distinguishability of the parameters themselves only (i.e., the identity property) and were restricted to  known coupling functions  \cite{Timme:14,Villaverde:14}.
 Our analysis characterizes necessary conditions to distinguish any property of the interaction matrix under uncertain coupling functions, and it can be straightforwardly extended to  include arbitrary-order interactions (e.g., $x_i x_j x_k$) and some nonlinear parametrizations (e.g., $x_i/(a_{ij} + x_j)$), SI-12. 
The analysis of uncertain coupling functions is motivated by existing NR algorithms that completely ignore our knowledge about the system  dynamics,  \cite{villaverde2014mider} and references therein. 
It is also possible to analyze the effect of noise and more general uncertainty of the coupling functions at the cost of less constructive results \cite{picci1977some}.
%We also point out that \textcolor{red}{our results are more constructive but less general than the analysis of identifiability provided in \cite{picci1977some}, allowing for example to characterize the separate contribution of the data and unceratin coupling function to the indistinguishability, and to determine that some useful properties cannot be reconstructed but the adjacency yes}.

Our results indicate that a better characterization of the uncertainty in the system's coupling functions and prior information of the interaction matrix are extremely useful to make practical improvements in network reconstructions.
This, in turn, calls for the
design of better algorithms (SI-4) that 
incorporate such information, and that  provide a guarantee of correct network
reconstruction.

%%%%%%%%%%%%%%%%%%%%%%%%
%%%%
%%%%
%%%%
%%%%
%%%%
%%%%%%%%%%%%%%%%%%%%%%%%

\begin{acknowledgments}
We thank Jean-Jacques Slotine and Travis Gibson for reading preliminary version of this paper. 
This work was supported by the CONACyT
postdoctoral grant 207609; 
Army Research Laboratories (ARL) Network
Science (NS) Collaborative Technology Alliance (CTA) grant: ARL NS-CTA
W911NF-09-2- 0053; 
DARPA Social Media in Strategic Communications
project under agreement number W911NF-12-C-002; 
the John Templeton
Foundation: Mathematical and Physical Sciences grant no. PFI-777; 
and the European Union grant no. FP7 317532 (MULTIPLEX).
\end{acknowledgments}

%\bibliographystyle{ieeetr}
\bibliography{NetworkReconstruction,NetworkReconstructionYang}

\end{document}